\documentclass[aps,prb,twocolumn,showpacs,amsmath,amssymb,amsfonts,floats,floatfix,balancelastpage,dvips,superscriptaddress]{revtex4}

\usepackage{epsfig} 
\usepackage{psfrag}
\usepackage{braket}
\usepackage{color}
\usepackage{graphicx}
\usepackage{hyperref}
\hypersetup{dvips,
  pdfauthor={Christopher N. Varney, Kai Sun, Marcos Rigol, and Victor Galitski},
  pdftitle={Interaction effects and quantum phase transitions in topological insulators},
  letterpaper,
  colorlinks=true,
  linkcolor=blue,
  citecolor=blue,
  pdfpagemode=UseNone
}
\usepackage{breakurl}

\newcommand{\bbraket}[1]{\braket{\hspace{-2pt}\braket{#1}\hspace{-2pt}}}

\begin{document}

\title{Interaction effects and quantum phase transitions in
  topological insulators}

\author{Christopher N. Varney}
\affiliation{Department of Physics, Georgetown University, Washington,
  DC 20057, USA}  
\affiliation{Joint Quantum Institute and Department of Physics, University of
  Maryland, College Park, Maryland 20742, USA}

\author{Kai Sun} 
\affiliation{Joint Quantum Institute and Department of Physics, University of
  Maryland, College Park, Maryland 20742, USA}
\affiliation{Condensed Matter Theory Center, Department of Physics, University
  of Maryland, College Park, Maryland 20742, USA}

\author{Marcos Rigol}
\affiliation{Department of Physics, Georgetown University, Washington,
  DC 20057, USA}

\author{Victor Galitski}
\affiliation{Joint Quantum Institute and Department of Physics, University of
  Maryland, College Park, Maryland 20742, USA}
\affiliation{Condensed Matter Theory Center, Department of Physics, University
  of Maryland, College Park, Maryland 20742, USA}

\begin{abstract}
  We study strong correlation effects in topological insulators via
  the Lanczos algorithm, which we utilize to calculate the exact
  many-particle ground-state wave function and its topological
  properties. We analyze the simple, non-interacting Haldane model on
  a honeycomb lattice with known topological properties and
  demonstrate that these properties are already evident in small
  clusters. Next, we consider interacting fermions by introducing
  repulsive nearest-neighbor interactions. A first-order quantum phase
  transition was discovered at finite interaction strength between the
  topological band insulator and a topologically trivial Mott
  insulating phase by use of the fidelity metric and the
  charge-density-wave structure factor. We construct the phase diagram
  at $T = 0$ as a function of the interaction strength and the complex
  phase for the next-nearest-neighbor hoppings. Finally, we consider
  the Haldane model with interacting hard-core bosons, where no
  evidence for a topological phase is observed. An important general
  conclusion of our work is that despite the intrinsic non-locality of
  topological phases their key topological properties manifest
  themselves already in small systems and therefore can be studied
  numerically via exact diagonalization and observed experimentally,
  {\it e.g.}, with trapped ions and cold atoms in optical lattices.
\end{abstract}


\maketitle

%%%%%%%%%%%%%%%%%%%%%%%%%%%%%%%%%%%%%%%%%%%%%%%%%%%%%%%%%%%%%%%%%%%%%%%%
\section{\label{sec:intro}Introduction}
%%%%%%%%%%%%%%%%%%%%%%%%%%%%%%%%%%%%%%%%%%%%%%%%%%%%%%%%%%%%%%%%%%%%%%%%
The conventional definition of a topological insulator (TI) relies on an
analysis of non-interacting band structures and from the mathematical point of
view represents topological classification of the spectrum of matrix
Hamiltonians.\cite{haldane1988,kane2005,bernevig2006,fu2007,
  moore2007,roy2009,qi2008a} While this approach is non-generalizable to
interacting systems, it is clear that they may exhibit physical phenomena
associated with non-trivial topological spectra. Qi and co-workers
recently suggested a more general definition of a topological insulator,
potentially suitable to interacting systems, which is defined as a state where
charged carriers give rise to axion electrodynamics with a non-trivial
electro-magnetic coupling term.\cite{qi2008a,li2010} One possible approach
would be to use the Volovik formula for topological indices, which involves
exact Green's functions.\cite{volovik2010} In addition, a technique utilizing
a Green's functions formula for time-reversal invariant topological insulators
was recently presented.~\cite{wang2010} Finally, the topological properties
can also be studied via explicit analysis of the edge states and the density
of states (DOS) spectrum with an eye on the localized boundary modes that
would connect bands across a gap.

Interacting topological models are expected to be even richer than the
non-interacting systems and may host a variety of different phases and phase
transitions. However, despite this expected rich variety of phenomena and
fundamental interest, the effect of interactions in topological insulators has
remained largely unexplored and the current knowledge of this issue is rather
limited.\cite{raghu2008,sun2009,pesin2010,dzero2010,rachel2010,wen2010} The
existing open questions include the following. (i)~Can topological properties
be tuned by interactions? (ii)~What is the nature of the quantum phase
transitions separating phases from different topological classes? (iii)~Are
topological Mott insulators possible, where the non-trivial phase would arise
entirely due to interactions? And an even more provocative question: (iv)~can
topological Mott insulators exist in bosonic systems? The main difficulty in
addressing these questions is that they all involve strong correlation
physics, where essentially no quantitatively reliable analytical methods
exist. The usefulness of quantum Monte Carlo
techniques\cite{blankenbecler1981} is expected to be limited by the infamous
sign problem.\cite{hirsch1985,loh1990} The last resort resides in exact
diagonalization methods and other unbiased numerical techniques, which are
limited to small system sizes as the dimension of the Hilbert space grows
exponentially with system size. The latter circumstance is important,
particularly because topological phases involve non-local effects and it is
not clear if they survive in small systems [{\em e.g.}, due to detrimental
effect of tunneling between the edge states in finite lattices with open
boundary conditions (OBC)].

In this paper, we employ the Lanczos algorithm\cite{lanczos1950, cullum1985}
to calculate the exact ground-state wave function of interacting lattice
models which are known or expected to have a topological ground state. We
focus our analysis on four observables that provide unique insights into the
properties of these lattice models. The topological phase is examined with the
local density of states (LDOS), which we determined for both open and periodic
boundary conditions (PBC). As we will discuss in detail below, the key
topological features are still discernible in relatively small clusters (at
least in two-dimensional lattices), which are within the reach of exact
diagonalization techniques. The next two observables of interest are the edge
currents and the structure factor, which we define carefully in the next
section. The final observable of interest is the ground-state fidelity metric
$g$, defined below in Eq.~\eqref{eqn:fidmet}. The fidelity metric is related
to the rate of change in the overlap between the ground-state wave function in
Hamiltonians that differ by a small control parameter. The ground-state
fidelity was originally studied in quantum information theory and has been
shown to be a sensitive indicator of quantum phase
transitions.\cite{zanardi2006,you2007,buonsante2007, chen2007,bodyfelt2007,
  campos2007,campos2008,garnerone2009,gu2009,rigol2009} In conjunction with
the structure factor, the fidelity metric is used to characterize the nature
of the transition from the topological insulator to the topologically trivial
Mott insulator.

To test the feasibility of the Lanczos algorithm and to explore the main
questions, we chose the Haldane model on a honeycomb lattice, but
with-nearest-neighbor repulsive interactions. The non-interacting base model
is known to have a simple topological phase transition to a trivial band
insulator,\cite{haldane1988} which can be tuned by adjusting a staggered
chemical potential, $M$. Moreover, in the limit of weak interaction strengths,
the bulk gap and non-trivial topology are known to
persist.\cite{niu1985,ishikawa1987,volovik1989} This allows us to calibrate
the techniques on the non-interacting and weakly interacting fermion model,
whose topological ground state is known and well understood.  The main
conclusion of this initial analysis is that the local density of states for
the bulk and the edge are the most useful metrics to probe the topological
nature of the ground state. On the other hand, we find that the existence of
the chiral edge currents may be misleading, as those may be present even in
topologically trivial phases and their origin and stability cannot be
determined just by looking at the current texture in the ground state. Indeed,
this result has been independently verified for non-interacting
systems.\cite{sonin2010}

\begin{figure}[t]
  \centering
  \includegraphics*[height=0.8\columnwidth,angle=-90,viewport=10 48 345 760]{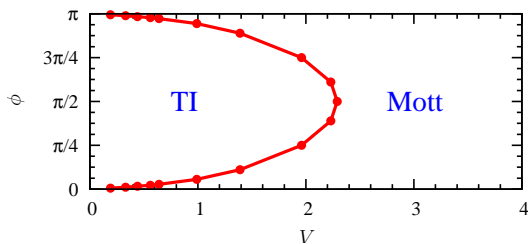}
  \caption{\label{fig:pd}
    (Color online) Schematic $V$-$\phi$ phase diagrams for spinless fermions
    in the Haldane model at half filling.  For $0 < \phi < \pi$, the system is
    a topological insulator (TI) at weak coupling and becomes a Mott insulator
    at finite interaction strengths via a first-order phase transition.  At
    $\phi = 0$ and $\phi = \pi$, the TI phase is not possible. Instead, the
    phase at weak coupling is a semimetal.  The points in the phase boundary
    were calculated for a 24-site cluster (24D).
  }
\end{figure}

With these caveats in mind, we analyze the strongly interacting Haldane model
with spinless fermions. The model is characterized by four independent
parameters: nearest-neighbor hopping amplitude, $t_1$, next-nearest-neighbor
hopping amplitude, $t_2$, and its complex phase $\phi$, and the
nearest-neighbor coupling, $V$. We limit our analysis of the interacting model
to the case of zero staggered potential, $M = 0$, which apart from the trivial
case of $\phi = 0$ (which is a semimetal) is guaranteed to be a topological
insulator for $V \ll t_{1,2}$. We find that as $V$ is increased, the system
undergoes a topological quantum phase transition into a topologically trivial
Mott insulator with the critical interaction strength $V_c(\phi) = V_c(\pi -
\phi)$ monotonically increasing increasing from $V_c(\pi / 180) = 0.19 t_1$ to
$V_c(\pi / 2) = 2.29 t_1$ within half a period (see Fig.~\ref{fig:pd}). The
appearance of the Mott insulator is not surprising, but it is not a priori
clear if the Mott transition must always coincide with the loss of the
topological order or if there are two separate transitions. Below, we show
compelling evidence that the former scenario is the one realized in this
model, that there is a single first-order quantum phase transition at
intermediate couplings.

Finally, we populated the Haldane honeycomb lattice with interacting hard-core
bosons in the hope that a phase with non-trivial topological properties may
emerge in this Bose-Hubbard model at intermediate couplings. However, it was
found not to be the case. At half filling, the system remains a trivial
superfluid until considerable values of nearest-neighbor repulsion parameter
is reached, at which point a superfluid-to-Mott-insulator transition takes
place. However, it is not a topological Mott insulator because the DOS shows
no evidence of an edge state crossing the Hubbard gap. Because this hard-core
boson model has interesting features beyond the focus of this paper, we defer
a full discussion of its properties for future work.\cite{varney2010}

\begin{figure}[t]
  \centering
  \includegraphics*[width=\columnwidth,viewport=0 300 612 492]{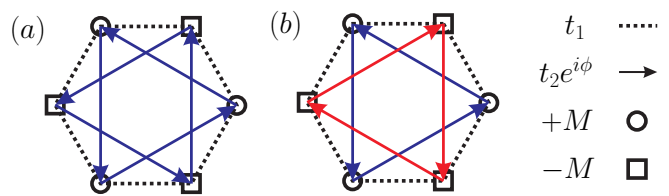}
  \caption{\label{fig:model}
    (Color online) Illustration of (a) the Haldane model and (b) the modified
    Haldane Hamiltonian. In (b), the direction of positive complex phase in
    the next-nearest-neighbor hopping on the sublattice denoted by the squares
    is reversed.
  }
\end{figure}

The remainder of this paper is structured as follows. In
Sec.~\ref{sec:model} we define the model and the key measurements used
to characterize the ground state: local density of states, structure
factor, and the fidelity metric. In addition, the Lanczos algorithm is
briefly discussed. In Sec.~\ref{sec:nonint}, we examine the exactly
solvable non-interacting Haldane model and show that topological
phases can be observed on small lattices. Next, we present the results
for interacting systems in Sec.~\ref{sec:int}, with
Sec.~\ref{subsec:fer} devoted to spinless fermions and
Sec.~\ref{subsec:hcb} to hard-core bosons. The main results and
outlook are summarized in Sec.~\ref{sec:summary}.

%%%%%%%%%%%%%%%%%%%%%%%%%%%%%%%%%%%%%%%%%%%%%%%%%%%%%%%%%%%%%%%%%%%%%%%%
\section{\label{sec:model}Model and Measurements}
%%%%%%%%%%%%%%%%%%%%%%%%%%%%%%%%%%%%%%%%%%%%%%%%%%%%%%%%%%%%%%%%%%%%%%%%
The Haldane Hamiltonian\cite{haldane1988} is a well-known model of
free fermions that features the anomalous quantum Hall state. In
real-space, the Hamiltonian is given by
\begin{align}
  \begin{aligned}
    H_{\rm Haldane} = &-t_1\sum_{\braket{i \, j}} \left( c^\dag_i c^{\phantom
      \dag}_j + c^\dag_j c^{\phantom \dag}_i \right)\\ &-
    t_2\sum_{\bbraket{i \, j}} \left(e^{i \phi_{ij}} c^\dag_i
    c^{\phantom \dag}_j + e^{-i \phi_{ij}} c^\dag_j c^{\phantom
      \dag}_i \right)\\ &+ M \sum_i (-1)^{\sigma(i)} n_i^{\phantom
      \dag} ,
  \end{aligned}
  \label{eqn:haldaneham}
\end{align}
where $c_i^\dag$ ($c_i^{\phantom\dag}$) represent the fermion
creation (annihilation) operators at site $i$ and $n_i =c^\dag_i
c_i^{\phantom\dag}$ is the corresponding number operator. Here $M$ is
a staggered potential that breaks the symmetry between the two
sublattices of a honeycomb lattice, indicated by odd or even values of
$\sigma(i)$, and $t_1$ ($t_2$) are the
nearest-neighbor (next-nearest-neighbor) hopping amplitudes. The
next-nearest-neighbor hopping term has a complex phase $\phi_{ij} =
\pm \phi$. In the Haldane model, the origin of this term is a varying
magnetic field that has zero net flux in a given hexagon. This is
illustrated in Fig.~\ref{fig:model}(a), where the sign of the complex
phase is positive in the direction of the vector between
next-nearest-neighbor sites. In Fig.~\ref{fig:model}(b), we show a
slightly modified version of the Hamiltonian, which differs from the
Haldane Hamiltonian in that the direction of positive complex phase on
one sublattice is reversed.

In this paper, we investigate the properties of the Haldane model for
interacting particles (spinless fermions or hard-core bosons) at
half filling. The Hamiltonian $H = H_{\rm Haldane} + H_{\rm int}$,
where
\begin{align}
  H_{\rm int} &= V \sum_{\braket{i \, j}} n_i^{\phantom \dag} n_j^{\phantom \dag}
\end{align}
and $V$ describes a repulsive nearest-neighbor interaction.

To study the interacting system, we utilize the Lanczos
algorithm.\cite{lanczos1950} (Details of the algorithm in the context of
Hermitian matrices can be found in Ref.~\onlinecite{cullum1985}.)  The main
advantage of this technique is that it provides extremely accurate information
about the ground-state wave function for interacting quantum
systems. Unfortunately, memory limitations restrict the size of the clusters
that can be studied. For the Hamiltonians of interest in this paper, we
consider various clusters with periodic boundary conditions. The largest
cluster studied has 30 sites, which at half filling has a Hilbert space of
$30!/(15!)^2\sim 1.55\times10^8$, near the limit of what is accessible on
present-day computers. We also consider one cluster with 24 sites and open
boundary conditions, which is used to study surface related effects. As
discussed above in Sec.\ \ref{sec:intro}, the ground-state wave function
determines the topological nature of a quantum system. More importantly, as we
will show in Sec.~\ref{sec:nonint}, the topological properties in the
anomalous quantum Hall system can be observed for small systems with only $24$
sites (or even smaller). These important features of the topological insulator
make the Lanczos algorithm a suitable tool in the study of the topological
properties.

As a vital feature of the Haldane model, there is no net flux in each unit
cell of the lattice. The absence of net flux turns out to be crucial in the
study of the topological edge states. As we implement the open boundary
condition, this could introduce a local flux near the boundary. This local
flux induces Faraday currents at the edge, which hybridize with the
topological edge currents and prevent a clear observation of the topological
edge states. Thus, it is crucial for the edge configuration to be chosen
carefully to avoid the net flux at the edge. In the numerical works reported
here, we removed the next-nearest-neighbor hopping terms between the sites on
the edge that do not pass through a hexagon to preserve this property.

One of the distinguishing characteristics of a topological insulator is the
presence of a (topologically protected) conducting edge state with an
insulating bulk. The LDOS can be measured via scanning tunneling microscopy in
conventional condensed matter systems.\cite{binnig1987} For a noninteracting
system, it is defined as
\begin{align}
  \label{eqn:ldos}
  N_i(\omega) = \sum_n |\braket{i|\Psi_n}|^2 \delta(\omega - E_n),
\end{align}
where $|\braket{i|\Psi_n}|^2$ is the weight of the state $\ket{\Psi_n}$ with a
particle in site $i$. In the Lanczos algorithm, the local density of states is
given by
\begin{widetext}
\begin{align}
  \label{eqn:lanczosdos}
  N_i(\omega) = \left\{
  \begin{array}{cc}
    \displaystyle\sum_n \left|\Braket{\Psi_n^{N-1}| c_i^{\phantom\dag} | \Psi_0^N}
    \right|^2 \delta \left[\omega - \left(E_n^{N-1} - E_0^N \right) \right] &
    \omega < \mu\\
    \displaystyle\sum_n \left|\Braket{\Psi_n^{N+1}| c_i^\dag | \Psi_0^N} \right|^2
    \delta\left[\omega + \left(E_n^{N+1} - E_0^N \right) \right] & \omega >
    \mu
  \end{array} \right. ,
\end{align}
\end{widetext}
where $\ket{\Psi_n^N} = \sum_m a_m^n \ket{\phi_m^N}$ is the $n$th eigenvector
with particle number $N$ and energy eigenvalue $E_n^N$ and $\ket{\phi_m^N}$
are the orthonormalized vectors determined in the Lanczos procedure. In
addition, it can easily be shown\cite{dagotto1994} that
\begin{align}
  \left|\Braket{\Psi_n^{N^\prime}| \widehat{O} | \Psi_0^N} \right|^2 &=
  |a_0^n|^2 \Braket{\Psi_0^N|\widehat{O}^\dag \widehat{O}|\Psi_0^N} ,
\end{align}
where $N^\prime = N \pm 1$. Further details on dynamical properties in Lanczos
can be found in Refs.~\onlinecite{gagliano1987,gagliano1988,dagotto1994}.

To better understand the edge states in our models, we considered the current
on every bond in the lattice. Between sites $m$ and $n$, the magnitude of the
density current $J_{mn}$ is defined to be\cite{rigol2008}
\begin{align}
  \label{eqn:current}
  J_{mn} &= \frac{i q r_{mn}}{\hbar} \left( t_{mn} c_m^\dag
    c_n^{\phantom \dag} - t_{nm} c_n^\dag c_m^{\phantom \dag} \right) ,
\end{align}
where $q$ is the charge carried by a particle and $r_{mn}$ is the magnitude of
the position vector between the two sites.

For large repulsive interaction strengths, the topological insulator will give
way to a trivial charge-density-wave (CDW) insulator through a topological
phase transition. In the limit $V \to\infty$, the ground state will be a
perfect CDW, where one of the two sublattices is occupied while the other is
empty, leaving lattice translational symmetry intact but breaking reflection
symmetry.\cite{wessel2007} The correlation function that describes the CDW
phase is
\begin{align}
  C({\bf r}_i - {\bf r}_j) &= \braket{ (n_i^a - n_i^b) (n_j^a - n_j^b) } ,
\end{align}
where $n_i^a$ and $n_i^b$ are the number operators on sublattice $a$ and $b$
in the $i$th unit cell, respectively. The corresponding structure factor is
\begin{align}
  \label{eqn:struct}
  S({\bf k}) &= \frac{1}{N}\sum_{i,j} e^{i {\bf k} \cdot
    ({\bf r}_i - {\bf r}_j)} C({\bf r}_i - {\bf r}_j) .
\end{align}
Because the long-range order is diagonal, $S({\bf k})$ will be maximal
at ${\bf k} = 0$ and we define $S_{\rm CDW} \equiv S({\bf k} = 0)$.

In order to study this quantum phase transition, we introduce an
observable related to the fidelity $F$. Let $\ket{\Psi_0(V)}$ be the
ground state of $H(V)$ and $\ket{\Psi_0(V + \delta V)}$ be the
ground state of $H(V + \delta V)$. The fidelity $F(V,\delta V)$ is
simply the overlap between these two wave functions
\begin{align}
  F(V,\delta V) &= |\braket{ \Psi_0(V) | \Psi_0(V + \delta V)}| .
\end{align}
The fidelity metric $g$ is a dimensionless, intensive quantity and can be
defined as
\begin{align}
  \label{eqn:fidmet}
  g(V,\delta V) &\equiv \frac{2}{N} \frac{1 - F(V,\delta V)}{(\delta
    V)^2} ,
\end{align}
where $N$ is the number of sites. For finite-size systems, a decrease in the
fidelity is a precursor to a quantum phase transition. This corresponds to a
peak in the fidelity metric, which should diverge in the limit $N \to \infty$
for a second-order phase transition. For a transition to the CDW phase, this
will coincide with a peak in $S({\bf k} = 0)$, which will similarly
diverge. The fidelity metric also tracks level crossings well. At a level
crossing, the wave functions $\ket{\Psi(V)}$ and $\ket{\Psi(V+\delta V)}$ will
be very different. Because the overlap between the two states is negligible,
the fidelity metric is sharply peaked. This, in conjunction with a jump in the
structure factor $S({\bf k} = 0)$, will distinguish a first-order phase
transition. In the calculations that follow, we take $\delta V = 10^{-4}$,
which is sufficiently small to ensure results consistent with $\delta V \to
0$.

%%%%%%%%%%%%%%%%%%%%%%%%%%%%%%%%%%%%%%%%%%%%%%%%%%%%%%%%%%%%%%%%%%%%%%%%
\section{\label{sec:nonint}Noninteracting Systems}
%%%%%%%%%%%%%%%%%%%%%%%%%%%%%%%%%%%%%%%%%%%%%%%%%%%%%%%%%%%%%%%%%%%%%%%%
\begin{figure}[t]
  \centering
  \includegraphics*[width=\columnwidth,viewport=0 0 612 754]{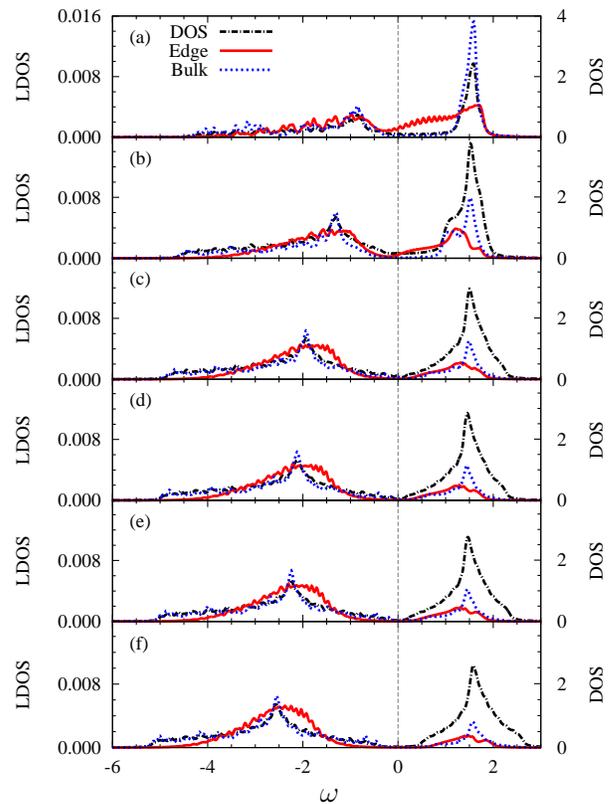}
  \caption{\label{fig:haldanedos}
    (Color online) Total density of states and local density of states in the
    Haldane model for a site on the edge and in the bulk for (a) $M = 0.0$,
    (b) $M = 0.5$, (c) $M = 1.0$, (d) $M = 1.1$, (e) $M = 1.2$, and (f) $M =
    1.5$. The calculations were performed on a $N = 420$-site cluster with
    $t_1 = 1.0$, $t_2 = 0.3$, and $\phi = \pi / 4$. The gap in the edge states
    opens at $M_c = 1.1$. The vertical, gray line indicates the chemical
    potential.
 }
\end{figure}

\begin{figure}[t]
  \centering
  \includegraphics*[width=\columnwidth,viewport=0 0 612 754]{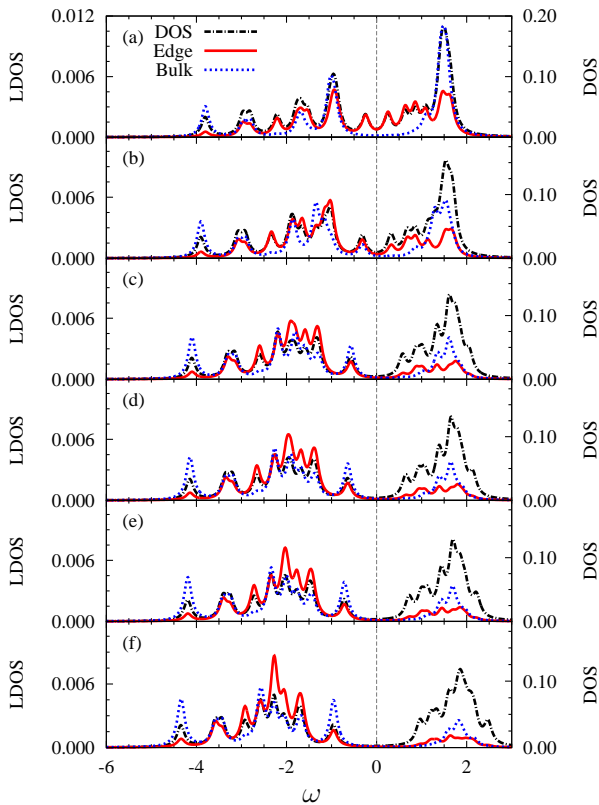}
  \caption{\label{fig:haldanedos2} 
    (Color online) Total density of states and local density of states in the
    Haldane model for a site on the edge and in the bulk for (a) $M = 0.0$,
    (b) $M = 0.5$, (c) $M = 1.0$, (d) $M = 1.1$, (e) $M = 1.2$, and (f) $M =
    1.5$. The calculations were performed on a $N = 24$-site cluster with the
    same parameters as in Fig.~\ref{fig:haldanedos}. The vertical, gray line
    indicates the chemical potential. Because of the finite size, the gap
    opens at smaller $M$ and is, in general, larger than in
    Fig.~\ref{fig:haldanedos}.
  }
\end{figure}

\begin{figure}[t]
  \centering
  \includegraphics*[width=\columnwidth,viewport=0 0 612 782]{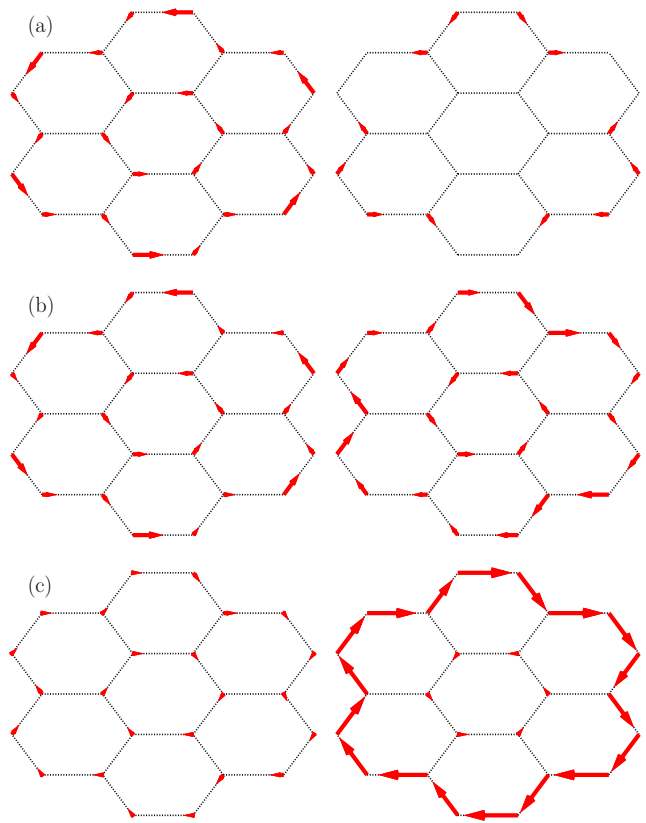}
  \caption{\label{fig:current} 
    (Color online) Nearest-neighbor currents for the Haldane (left) and
    modified Haldane (right) Hamiltonians on a $N = 24$-site cluster and
    parameters $t_1 = 1.0$, $t_2 = 0.3$, and $\phi = \pi / 4$ with (a) $M =
    0$, (b) $M = 0.10$, and (c) $M = 1.5$. In the modified Haldane model, a
    non-zero circulating current exists for $M \ne 0$, i.e., the moment
    chiral symmetry is broken.
  }
\end{figure}

\begin{figure}[t]
  \centering
  \includegraphics*[width=\columnwidth,viewport=0 0 612 754]{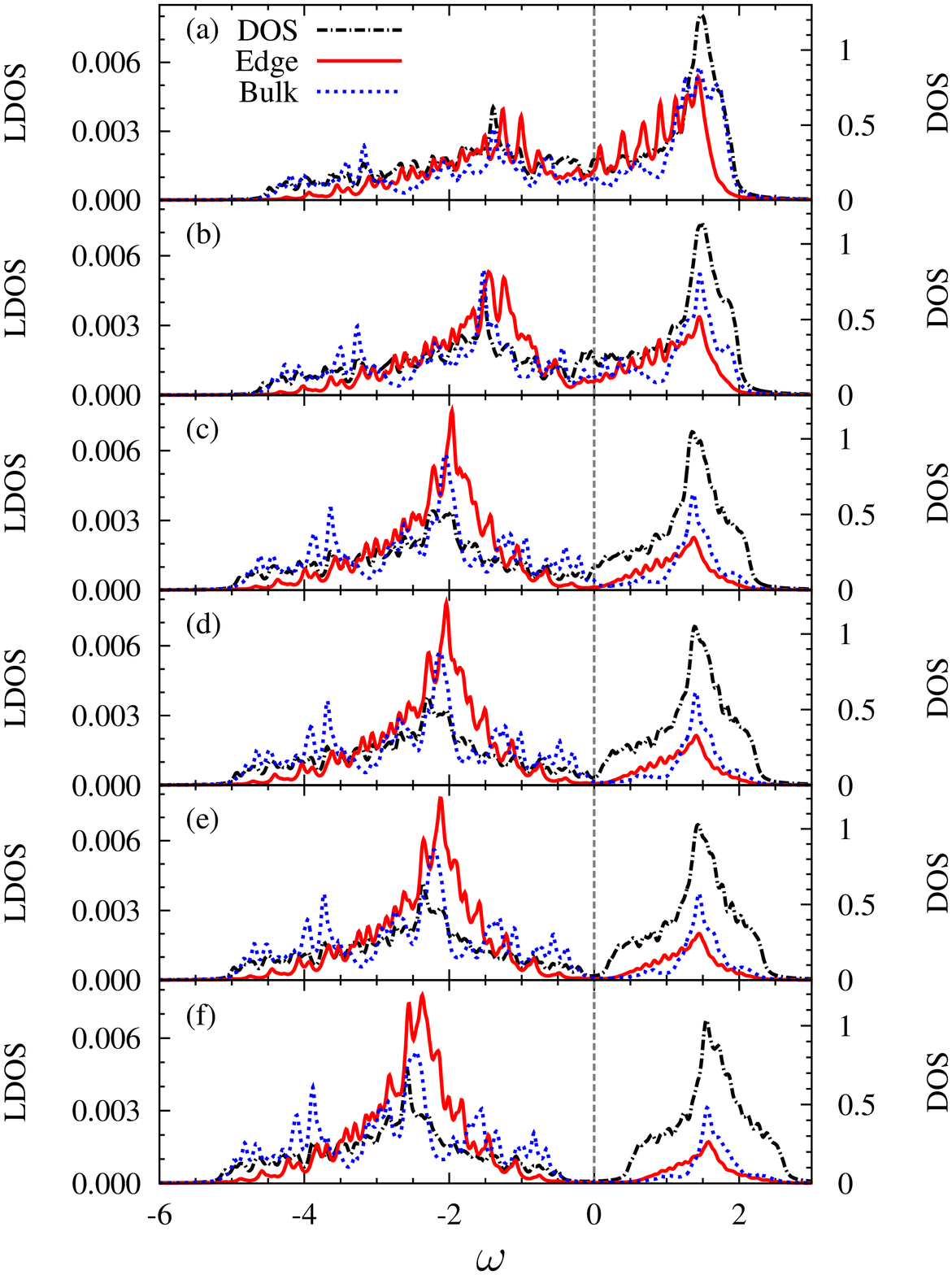}
  \caption{\label{fig:mhaldanedos} 
    (Color online) Total density of states and local density of states in the
    modified Haldane model for a site on the edge and in the bulk for (a) $M =
    0.0$, (b) $M = 0.5$, (c) $M = 1.0$, (d) $M = 1.1$, (e) $M = 1.2$, and (f)
    $M = 1.5$. The calculations were performed on a $N = 420$-site cluster
    with $t_1 = 1.0$, $t_2 = 0.3$, and $\phi = \pi / 4$. The gap in the edge
    states opens at $M_c = 1.1$. The vertical, gray line indicates the
    chemical potential.
 }
\end{figure}

We first study the noninteracting fermions, which are exactly solvable, to
compare our numerical results with the known solution. In addition, this limit
allows for the study of large clusters and we can carefully address whether
the small clusters that are accessible with the Lanczos algorithm are suitable
for observation of a topological insulator.

As Haldane was able to show previously,\cite{haldane1988} there is a phase
transition between the topological insulator and the topologically trivial
charge insulator as one increases the value of $M$. For $M$ below the critical
value of $M_c = 3^{3/2} t_2 \sin \phi$, the system is a topological insulator
with an insulating bulk and a chiral metallic edge. As $M$ approaches $M_c$,
the insulating gap in the bulk decreases and closes at $M = M_c$ while the
system becomes a semimetal. If one further increases $M$, an insulating gap
develops in both the bulk and the edge. However, this insulator is
topologically trivial and there are no gapless edge states with topological
protection.

Unlike an ordinary phase transition, this transition was beyond the framework
of the Landau's theory of phase transitions. The two phases discussed above
cannot be distinguished by any local parameter and no symmetry was broken
across this transition. As a result, Landau's free energy cannot be
defined. The only distinction between these two phases are their topological
property, which is described by a topological index known as the Chern
number.\cite{haldane1988,sun2008,cheng2010} Experimentally, one can
distinguish these two phases by their edge properties. We also emphasize here
that this phase transition shows no discontinuity for any thermodynamical
quantity and a similar example of a topological transition is the transition
between the strong and weak pairing in $p_x+ i p_y$ superconductors (at least
within the mean-field treatment).\cite{read2000,cheng2010}

This topological transition can be observed by measuring the local density of
states [Eq.~\eqref{eqn:ldos}]. Since we deal here with finite-size systems,
the energy spectrum is discrete. To compensate for this, we have broadened the
delta function peaks with a Lorentzian of width $\gamma = 0.01t_1$.  As shown
in Fig.~\ref{fig:haldanedos}, for small $M$, the LDOS has no energy gap for a
site on the edge, indicating clearly a metallic edge. However, for the bulk
site, the same spectrum becomes gapped, reflecting an insulating bulk.

To estimate the effects of finite-sized systems on this transition, we reduced
the size of the cluster down to 24 sites (12 unit cells). In
Fig.~\ref{fig:haldanedos2}, the bulk gap and the gapless edge states can still
be observed, indicating that the topological property of the anomalous quantum
Hall state is robust with regard to finite-sized clusters. A similar effect
was also observed in a system fundamentally different from the model studied
here. As reported in Ref. \onlinecite{chen2009}, the topological ordering in
the Kitaev model\cite{kitaev2006} was observed for small system sizes. While
the topological ordering found by Chen {\em et al.} has little direct relation
with the results presented here, the fact that the same phenomenon is observed
in these two systems suggests that robustness against finite-size effects
might be a common property shared by different topological states of matter.

We also computed the nearest-neighbor current [Eq.~\eqref{eqn:current}] in the
system, which is shown for $M = 0$, $0.1$, and $1.5$ in Fig.~\ref{fig:current}
(next-nearest-neighbor currents are not shown because they are an artifact of
the complex hopping in the Hamiltonian and exist for all parameters). Note
that the topological state has a clear chiral edge current (rotating in the
clockwise direction), while the current in the bulk is negligible. However, in
the topologically trivial insulator with $M > M_c$, the chiral current remains
until $M$ becomes large. This is because Haldane's model explicitly breaks
chiral symmetry, which allows for a rotating edge current. Due to this effect,
by measuring only the edge current, it is not sufficient to distinguish the
topologically trivial insulator from the topological insulator.  Indeed, the
proper means of distinguishing the topological state is via the LDOS.

\begin{figure*}[t]
  \centering
  \includegraphics*[width=\textwidth,viewport=00 160 612 611]{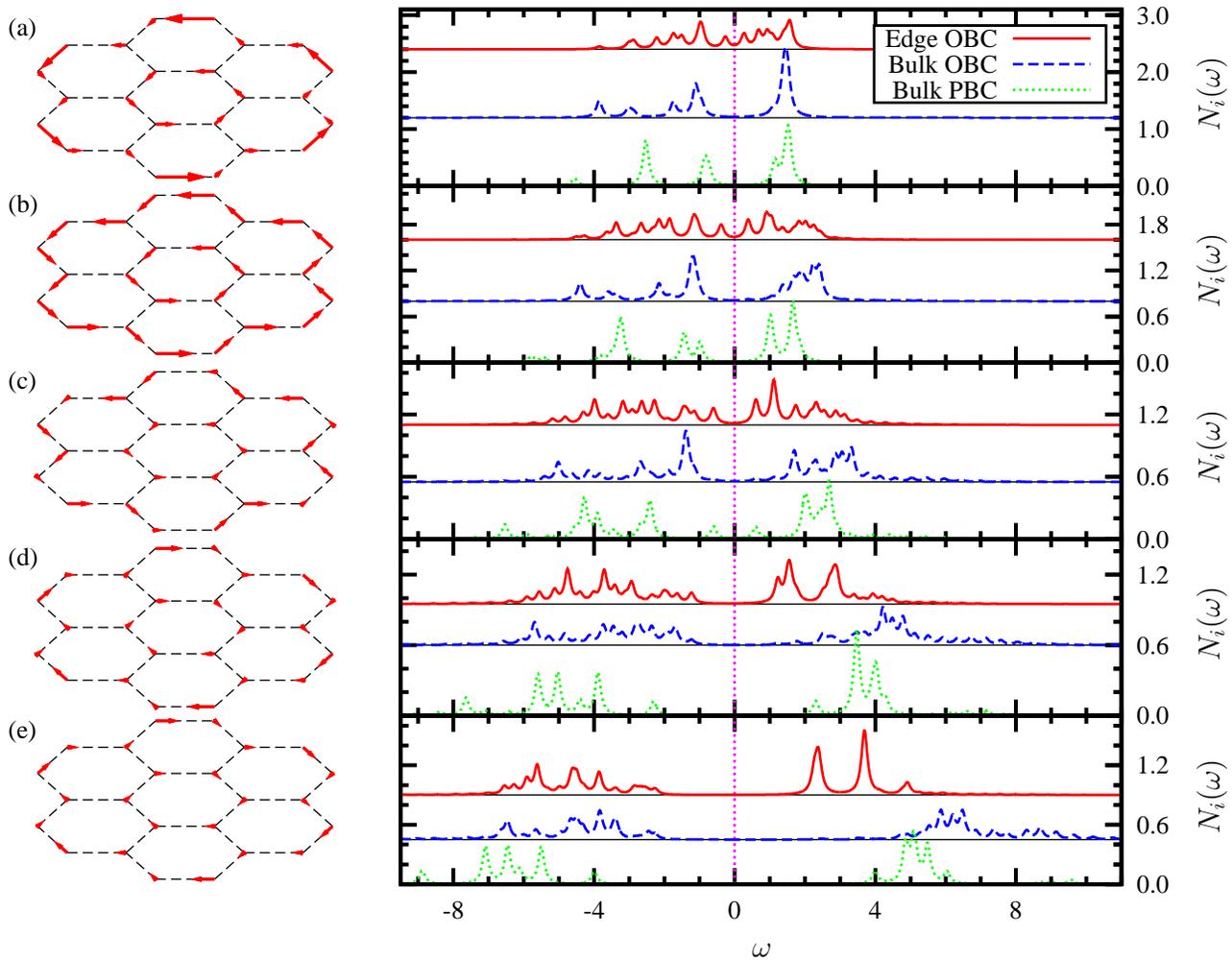}
  \caption{\label{fig:Fcurrdos} 
    (Color online) Nearest-neighbor currents (left) for spinless
    fermions on a 24-site cluster with OBC, $t_1 = 1.0$,
    $t_2 = 0.3$, $\phi = \pi / 4$, and interaction strengths (a) $V =
    0$, (b) $V = 1$, (c) $V = 2$, (d) $V = 3$, and (e) $V = 4$. Local
    density of states (right) for the edge (OBC) and the bulk (shown
    for both OBC and PBC). The zeros of $N_{\rm
      edge}(\omega)$ and $N_\text{bulk, OBC}(\omega)$ are shifted
    upward for clarity, and the frequency $\omega$ is shifted so that
    the chemical potential is at $\omega = 0$ (which is also indicated
    by the dashed vertical line). The delta function is broadened by a
    Lorentzian of width $\gamma = 0.01t_1$.
%    , and local density of states (right) for the edge and the bulk
%    for spinless fermions on a 24-site cluster with open boundary
%    conditions, $t_1 = 1.0$, $t_2 = 0.3$, $\phi = \pi / 4$, and
%    interaction strengths (a) $V = 0$, (b) $V = 1$, (c) $V = 2$, (d)
%    $V = 3$, and (e) $V = 4$. Here we show the bulk LDOS for both open
%    (OBC) and periodic boundary conditions (PBC). The zeros of $N_{\rm
%      edge}(\omega)$ and $N_\text{bulk, OBC}(\omega)$ are shifted
%    upwards for clarity, and the frequency $\omega$ is shifted so that
%    the chemical potential is at $\omega = 0$ (which is also indicated
%    by the dashed vertical line). The delta function is broadened by a
%    Lorentzian of width $\gamma = 0.01t_1$.
  }
\end{figure*}

For comparison purposes, we also studied a modification to Haldane's model
[Fig.~\ref{fig:model}(b)], where the two sublattices have opposite flux
patterns. This model is topologically trivial in all of the parameter
regimes. For $M < M_c$, the system is a conductor (a Fermi liquid) at half
filling but becomes a topologically trivial insulator for $M > M_c$. At the
transition point $M = M_c$ (which is identical to the transition point in
Haldane's model), the system is a semimetal. This transition was observed
clearly in the LDOS (Fig.~\ref{fig:mhaldanedos}). In particular, the
insulating phase with $M > M_c$ was found to be gapped both in the bulk and on
the edge, which, as expected, signifies a topologically trivial
insulator. Moreover, the same behavior can be found even for the small
clusters with only 24 sites (not shown).

Although the modified Haldane's model is topologically trivial, this system
has its own significance in the study of the anomalous Hall effect. As in the
Haldane's model described previously, this model also has no net magnetic
flux. However, the flux pattern breaks the chiral and time-reversal symmetries
for any $M \ne 0$, even in the absence of the a net magnetic flux. These
symmetry properties are in close analogy to the (non-quantized) Hall effect,
although the magnetic field is zero after averaging over each unit cell in
this case. Due to this similarity, the conducting phase in the modified
Haldane's model with $0 < M < M_c$ is referred to as an anomalous Hall
state.\cite{haldane2004,sun2008} This anomalous Hall effect can be easily
observed by looking at the currents in the system (see panels on right side of
Fig.~\ref{fig:current}). At $M = 0$, the chiral symmetry is preserved, and the
current on the left and right edges flows in opposite direction in order to
preserve the chiral symmetry. However, for $M > 0$, the ground state of the
system is the metallic anomalous Hall state, which breaks the chiral
symmetry. Now, the currents at the opposite edge will not cancel each
other. Instead, they flow in the same direction and form a rotating current
circling around the system. We emphasize here that this edge current is not
related with a nontrivial topological structure in the ground state wave
function, and it only reflects the broken chiral
symmetry.\cite{haldane2004,sun2008,nagaosa2010}

%%%%%%%%%%%%%%%%%%%%%%%%%%%%%%%%%%%%%%%%%%%%%%%%%%%%%%%%%%%%%%%%%%%%%%%%
\section{\label{sec:int}Interacting Systems}
%%%%%%%%%%%%%%%%%%%%%%%%%%%%%%%%%%%%%%%%%%%%%%%%%%%%%%%%%%%%%%%%%%%%%%%%
\subsection{\label{subsec:fer}Spinless fermions}
Having shown that the topological insulator phase can be distinguished on
small lattices, we now consider the properties of interacting systems. In this
section, we consider spinless fermions with repulsive nearest-neighbor
interactions (hard-core bosons are discussed in Sec.~\ref{subsec:hcb}). In the
discussion that follows, we present detailed results for the parameters $t_1 =
1.0$, $t_2 = 0.3$, $\phi = \pi / 4$, and $M = 0$ before discussing the results
for general $\phi$.

At the non-interacting limit, we have shown above that, at $M = 0$, the system
is a topological insulator and the two sublattices have the same occupation
number. In the strong-coupling limit $V\rightarrow \infty$, however, the
particles will all go to one of the two sublattices to avoid the energy
penalty for occupying two neighboring sites. At half filling, one sublattice
is fully filled while the other one is empty, resulting in a CDW insulating
phase. Note with the formation of the CDW, translational symmetry remains but
inversion symmetry was spontaneously broken. Hence, we conclude that the
system goes though a quantum phase transition as $V$ is increased. More
interestingly, the insulating phase in the weak-coupling limit is signified by
a nontrivial Chern number.\cite{haldane1988,sun2008,cheng2010} However, the
CDW phase in the strong-coupling limit has a trivial Chern number ($C = 0$)
and is topologically trivial. Therefore, the topology of the ground-state wave
function changes from nontrivial to trivial as $V$ increases. This change in
topology is referred as a topological transition. Within the current knowledge
of the topological insulators, it is not clear whether the quantum phase
transition and the topological transition are, in general, two separate
transitions or if they coincide to become a single phase transition.

We first consider the topological transition by examining the LDOS at the edge
and in the bulk. In Fig.~\ref{fig:Fcurrdos}, we show local currents (left) and
the local density of states (right) on a 24-site cluster with open boundary
conditions for several interaction strengths. At $V = 0$
[Fig.~\ref{fig:Fcurrdos}(a)], the system is a topological insulator, evidenced
by the bulk insulating gap and the metallic local density of states on the
edge. The circulating current on the edge is also consistent with the quantum
Hall state. As the interaction strength is increased, the gap in $N_{\rm
  bulk}(\omega)$ closes while the states near the chemical potential in
$N_{\rm edge}(\omega)$ are depleted. For $V = 2$, $3$, and $4$
[Fig.~\ref{fig:Fcurrdos}(c)-\ref{fig:Fcurrdos}(e)], the edge states are fully
gapped and the system is a trivial insulator. The local density of states in
this phase is characterized by a near complete depletion of the bulk states
closest to the gap in the upper band, effectively resulting in a larger
insulating gap in the bulk than on the edge. The topological transition takes
place at $V_T \approx 2$. Finally, we note that the trivial insulator at
intermediate coupling is characterized by small but non-zero edge currents
that travel in the opposite direction from the edge currents in the
topological insulating phase. In the limit of $V \gg t_1$, the edge currents
vanish.

\begin{figure}[t]
  \centering
  \includegraphics*[height=\columnwidth,angle=-90,viewport=0 0 612 792]{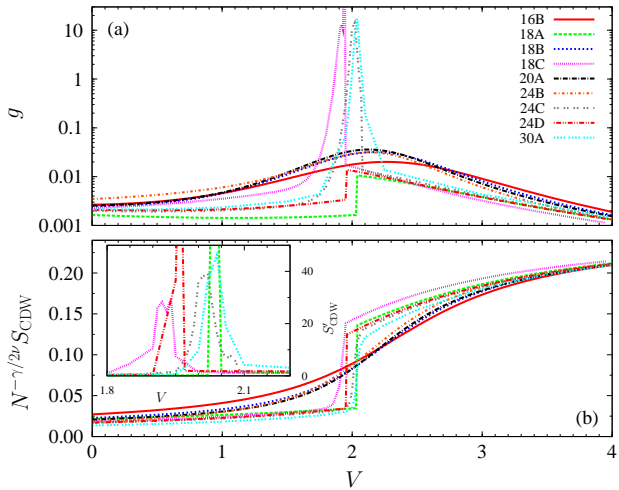}
  \caption{\label{fig:Ffidelity} 
    (Color online) (a) Fidelity metric $g$ as a function of interaction
    strength for various clusters with parameters $t_1 = 1.0$, $t_2 = 0.3$,
    $\phi = \pi / 4$, and $\delta V = 10^{-4}$. (b) Scaled CDW structure
    factor $N^{-\gamma / 2 \nu} S_{\rm CDW}$ for the same parameters (shown
    for $\eta = 0$ and $\gamma / \nu = 2$). All clusters that contain the zone
    corner (${\bf k} = K$) as a valid $k$ point exhibit a sharp, first-order
    transition at $V_c \approx 2.0$. Without this $k$ point, a smooth,
    continuous transition is observed. The inset shows the first derivative of
    the structure factor. The discontinuity (i.e., the peak) marks the
    transition point.
  }
\end{figure}

Next, we consider the nature of the quantum phase transition which leads to
the spontaneous breaking of inversion symmetry. Here, the momentum space
properties of the cluster are vital to the type of transition observed. The
noninteracting band structure for this Hamiltonian features a Dirac cone at
the zone corner $K = (2\pi/3, 2\pi/\sqrt{3})$. For an infinite system, this
will always be a valid momentum point, whereas for small clusters this is not
generally the case. Because the surface states in the band structure pass
through this point in reciprocal space, the order of the transition has a
clear dependence on the choice of clusters. This is reflected in the fidelity
metric and the scaled CDW structure factor $N^{-\gamma / 2 \nu} S_{\rm CDW}$,
which are shown in Fig.~\ref{fig:Ffidelity} for clusters from $N = 16$ to $N =
30$ sites with periodic boundary conditions (see the Appendix for more details
on the cluster shapes). If one studies clusters with a reciprocal space that
does not contain the zone corner, then a broad maximum in the fidelity metric
and a crossing of the scaled structure factor is observed at $V \approx
2.0$. On just this information, one would conclude that there is a
second-order phase transition at this interaction strength. However, clusters
that contain the $K$ point (18A, 18C, 24C, 24D, and 30A) exhibit markedly
different behavior. Instead, we find that the fidelity metric exhibits a very
large and sharp peak (18C, 24C, and 30A) or discontinuity (18A and 24D) at $V
\approx 2.0$. Furthermore, the CDW structure factor exhibits a jump at this
interaction strength, indicating a first-order phase transition. Indeed, when
we examine the first derivative of the structure factor [see
Fig.~\ref{fig:Ffidelity}(b) inset], there is removable discontinuity at $V_c
\approx 2.0$, which indicates that the topological transition and the quantum
phase transition occur at the same interaction strength.

\begin{figure}[t]
  \centering
  \includegraphics*[height=\columnwidth,angle=-90,viewport=0 0 612 792]{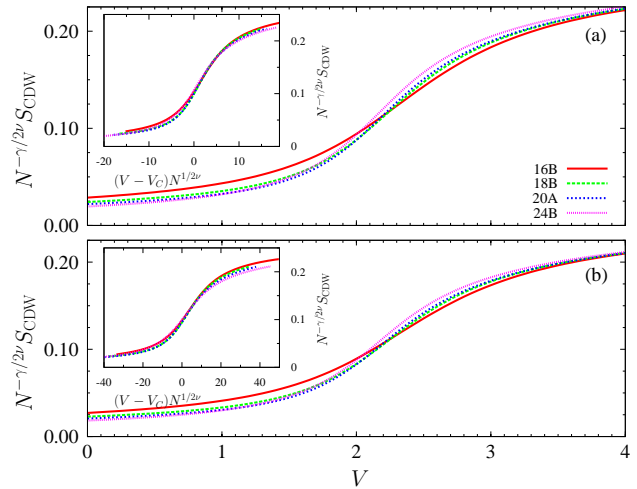}
  \caption{\label{fig:Fscaling} 
    (Color online) Scaled structure factor $N^{-\gamma / 2 \nu} S_{\rm CDW}$
    for (a) $z = 1$ and $\gamma / \nu = 1.96$ ($\eta = 0.04$) and (b) $z = 2$
    and $\gamma / \nu = 2$ ($\eta = 0$). In both cases, the transition point
    is found to be $V_c = 2.1 \pm 0.1$. In the insets, the horizontal axis is
    scaled as well, leading to the collapse of all data points into a single
    curve.
  }
\end{figure}

\begin{figure*}[t]
  \centering
  \includegraphics*[width=\textwidth,viewport=00 160 612 612]{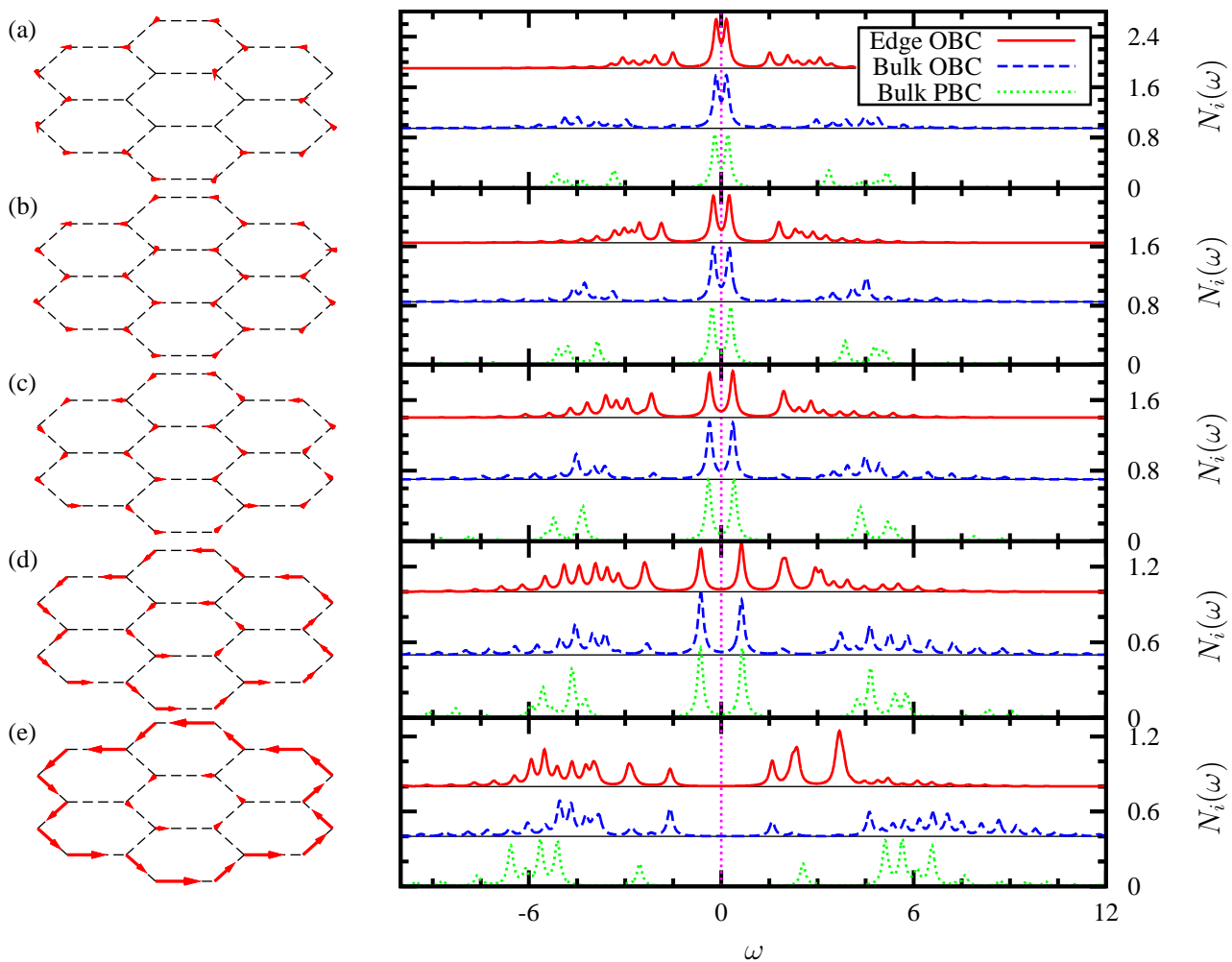}
  \caption{\label{fig:Bcurrdos}
    (Color online) Nearest-neighbor currents (left) for hard-core
    bosons on a 24-site cluster with OBC, $t_1 = 1.0$, $t_2 = 0.3$, 
    $\phi = \pi / 4$, and interaction strengths (a) $V = 0$, (b) $V =
    1$, (c) $V = 2$, (d) $V = 3$, and (e) $V = 4$. Local density of
    states (right) for the edge (OBC) and the bulk (shown for both OBC
    and PBC). The zeros of $N_{\rm edge}(\omega)$ and $N_\text{bulk,
      OBC}(\omega)$ are shifted upward for clarity, and the frequency
    $\omega$ is shifted so that the chemical potential (dashed
    vertical line) is at $\omega = 0$.
%    (Color online) Nearest-neighbor currents (left), and local density
%    of states (right) for the edge and the bulk for hard-core bosons
%    on a 24 site cluster with open boundary conditions, $t_1 = 1.0$,
%    $t_2 = 0.3$, $\phi = \pi / 4$, and interaction strengths (a) $V =
%    0$, (b) $V = 1$, (c) $V = 2$, (d) $V = 3$, and (e) $V = 4$. Here
%    we show the bulk LDOS for both open (OBC) and periodic boundary
%    conditions (PBC). The zeros of $N_{\rm edge}(\omega)$ and
%    $N_\text{bulk, OBC}(\omega)$ are shifted upward for clarity, and
%    the frequency is shifted so that the chemical potential (dashed
%    vertical line) is at $\omega = 0$.
  }
\end{figure*}

For the clusters that do not contain the $K$ point, the quantum phase
transition is a conventional one (in contrast to the topological phase
transitions discussed above). Because the inversion operator and the identity
operator form a $Z_2$ group, the symmetry-breaking pattern of this quantum
phase transition belongs to the $Z_2$ (also known as Ising) universality
class.  If the transition is second-order, it is then expected to follow the
scaling relation of the Ising model in $d + z$ dimensions. Here $d = 2$ is the
spatial dimension and $z$ is the dynamic critical exponent, which is usually a
positive integer indicating how many spatial dimensions the time dimension
shall be counted as.\cite{sachdev1999} For $z = 1$, the critical point follows
the three-dimensional (3D) Ising scaling, for $z = 2$ and above, the exponents
are the mean-field exponents.\cite{cowan2005,pathria1996}

We expect that the structure factor will scale according the following
rule
\begin{align}
  \label{eqn:scaling}
   N^{-\gamma / 2\nu} S_{\rm CDW} = f[(V - V_c)N^{1/2\nu}] \,,
\end{align}
where $N$ is the number of sites and $\gamma = \nu (2 - \eta)$. For a 3D Ising
model,\cite{cowan2005} (corresponding to the case of $z = 1$), $\nu = 0.7$ and
$\eta = 0.04$, while in four dimensions (4D) and
above\cite{pathria1996,lundow2009} ($z \ge 2$), $\nu = 0.5$ and $\eta = 0$. As
the proper value of $z$ is still an open question, we show the rescaled
structure factor $N^{-\gamma / 2\nu} S_{\rm CDW}$ as a function of interaction
strength for both of these cases in Fig.~\ref{fig:Fscaling}. If the transition
is second order, the set of curves will converge at the quantum critical point
$V = V_c$. Here, the critical point obtained in both cases remains almost
unchanged. Moreover, we show the universal scaling relation,
Eq.~\eqref{eqn:scaling}, in the insets of Fig.~\ref{fig:Fscaling} and observe
the same qualitative behavior for both $z = 1$ and $z = 2$. As there is no
discernible qualitative difference between the scaling with $z = 1$ and $z =
2$, it is clear that higher order corrections in the finite-size scaling are
relevant. For the largest system size shown in Fig.~\ref{fig:Fscaling}, the
linear system size is $L = N^{1/2} \approx 4.9$. Thus, significantly larger
system sizes need to be studied in order to properly determine the value of
$z$.

\begin{figure*}[t]
  \centering
  \includegraphics*[width=\textwidth,viewport=0 256 600 532]{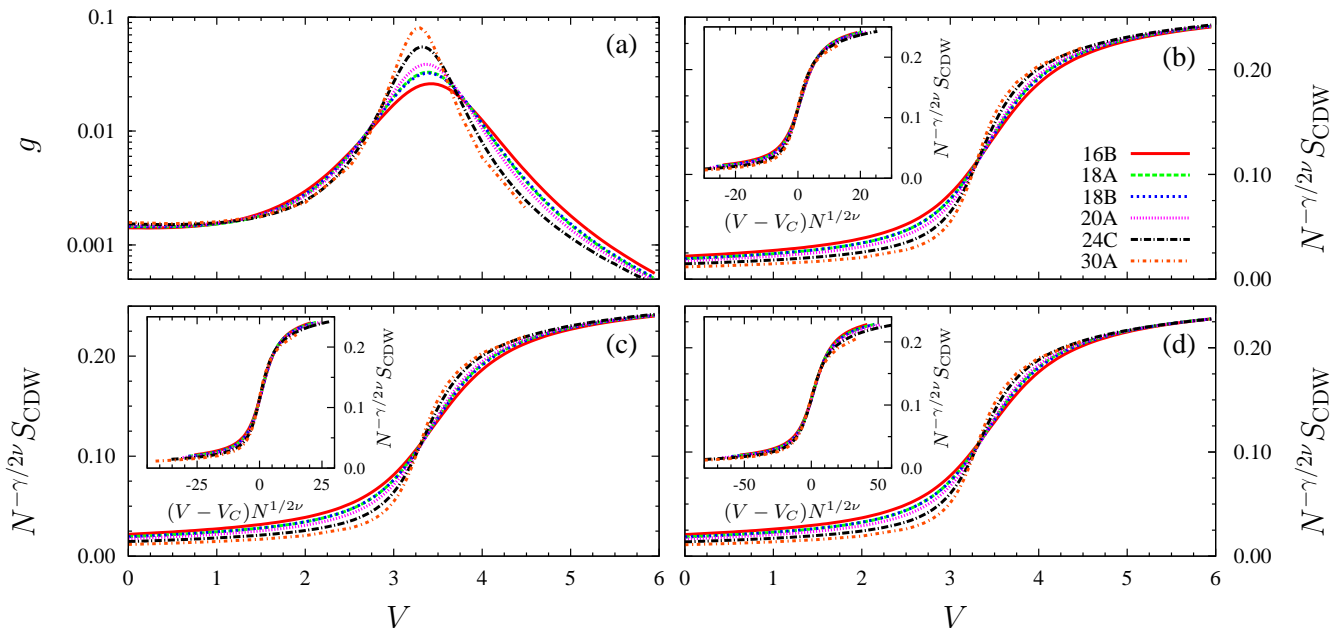}
  \caption{\label{fig:Bfidelity}
    (Color online) (a) Fidelity metric $g$ as a function of interaction
    strength for hard-core bosons in various clusters with parameters $t_2 =
    0.3$, $\phi = \pi / 4$, and $\delta V = 10^{-4}$. [(b)-(d)] Scaled CDW
    structure factor $N^{-\gamma / 2 \nu} S_{\rm CDW}$ for (b) 3D Ising ($z = 1$),
    (c) 3D XY ($z = 1$), and (d) 4D Ising / XY ($z = 2$). All of the curves
    intersect at the critical point $V_c = 3.27 \pm 0.05$ in (b) and (c) and
    $V_c = 3.28 \pm 0.05$ in (d).  In the insets of (b)-(d), the horizontal axis
    is also rescaled and all of the data points collapse into a single curve. 
  }
\end{figure*}

In Fig.~\ref{fig:pd}(a), we present a schematic phase diagram for this model
in the $V$-$\phi$ plane. Note that the phase boundary shown in the figure is
determined for a 24-site cluster from the maximum in the derivative of $S_{\rm
  CDW}$ [see Fig.~\ref{fig:Ffidelity}(b) inset]. At $\phi = 0$, there is no
magnetic field in the system to break the time-reversal symmetry and, at
weak coupling, the system is a semimetal, not a topological insulator. Once
time-reversal symmetry is broken ($0 < \phi < \pi$), a topological insulator
is observed for small $V$. There is a topological transition from the
topological insulator, and this coincides with a first-order quantum phase
transition to a topologically trivial Mott insulator. Furthermore, we note
that the topological phase is strongest at $\phi = \pi / 2$.
% The topological insulator undergoes a first-order quantum phase transition
% to a topologically trivial Mott insulator, and the topological phase is
% strongest at $\phi = \pi / 4$.

%%%%%%%%%%%%%%%%%%%%%%%%%%%%%%%%%%%%%%%%%%%%%%%%%%%%%%%%%%%%%%%%%%%%%%%%
\subsection{\label{subsec:hcb}Hard-core bosons}
%%%%%%%%%%%%%%%%%%%%%%%%%%%%%%%%%%%%%%%%%%%%%%%%%%%%%%%%%%%%%%%%%%%%%%%%
Next, we examined the properties of the Haldane model with hard-core bosons at
half filling and nearest-neighbor repulsive interactions, to check whether
there are phases with non-trivial topological properties. We are motivated by
the fact that in many instances hard-core bosons and spinless fermions are
known to have similar properties. In one dimension, one can use the Jordan-
Wigner transformation to map one onto the other, i.e., there is a one-to-one
correspondence between them. In higher dimensions, one can also expect some
similarities. For example, a simple band-structure calculation shows that, at
half filling, the ground state of noninteracting fermions on a hypercubic
lattice in the presence of a staggered potential $M$, like the one in
Eq.~\eqref{eqn:haldaneham}, is metallic for $M = 0$ and becomes insulating for
any nonzero value of $M$. For hard-core bosons, on the other hand, one has a
superfluid (Bose-Einstein condensed) phase for $M < M_c$ and an insulator
phase for $M>M_c$,\cite{hen09,hen10} i.e., both models have equivalent
ground-state phases. The only difference in this case is that for spinless
fermions the metal-insulator transition always occurs at $M_c = 0$, while for
hard-core bosons the superfluid-insulator transition occurs at a finite value
of $M_c$, which depends on the dimensionality of the system ($M_c = 0$ in one
dimension, as expected from the exact mapping). Hence, similar physics as
discussed in Sec.~\ref{subsec:fer} for spinless fermions would not be
unexpected in hard- core boson systems, where, of course, equivalent phases
may just appear for different ranges of values of the Hamiltonian parameters.

However, in Fig.~\ref{fig:Bcurrdos}, we show that this is not the case. For
small interaction strengths, the system is a superfluid with essentially no
edge current. As $V$ is increased, the system becomes an insulator
simultaneously on the edge and in the bulk, indicating a direct transition to
the topologically trivial Mott insulator. This Mott insulating state, like the
spinless fermion model, exhibits a circulating current on the edge and a
depletion of the states in the bulk of the upper band at intermediate coupling
strengths. Here the edge currents in the insulating state are larger than for
spinless fermions and consequently vanish at larger interaction strengths.

To determine the nature of the phase transition to the CDW phase we examined
the fidelity metric and CDW structure factor for various system sizes. The
peak in the fidelity metric [Fig.~\ref{fig:Bfidelity}(a)] grows with the
system size, which is an indicator of a second-order phase transition. Unlike
the spinless fermion case in Sec.~\ref{subsec:fer}, the universality class of
the transition is unclear. We have performed scaling analysis of the structure
factor for both Ising and XY universality classes with dimension $d + z$. The
critical exponents for Ising and XY scaling are similar in 3D and identical in
4D and above (the upper critical dimension of both models is $4$). Because
these exponents are similar and finite-size effects are present in our
calculations, we cannot determine which universality class, if any, is the
appropriate one. To illustrate this, we show the structure factor rescaled by
a factor of $N^{-\gamma / 2\nu}$ in
Figs.~\ref{fig:Bfidelity}(b)-\ref{fig:Bfidelity}(d) for 3D Ising, 3D
XY,\cite{hasenbusch1999} and 4D Ising/XY, respectively. There is no
discernible difference between the three cases. The curves in $N^{-\gamma / 2
  \nu} S_{\rm CDW}$ versus $V$ for all three cases cross at roughly the same
interaction strength ($V_c = 3.27 \pm 0.05$ in 3D and $V_c = 3.28 \pm 0.05$ in
4D) and the crossing point coincides with the peak in the fidelity
metric. Moreover, when plotting $N^{-\gamma / 2\nu} S_{\rm CDW}$ versus $(V -
V_c) N^{1/2\nu}$ (see Fig.~\ref{fig:Bfidelity} insets), all the resulting
curves almost lie on top of each other. A more detailed study of this model
will presented in future work.\cite{varney2010}

%%%%%%%%%%%%%%%%%%%%%%%%%%%%%%%%%%%%%%%%%%%%%%%%%%%%%%%%%%%%%%%%%%%%%%%%
\section{\label{sec:summary}Summary}
%%%%%%%%%%%%%%%%%%%%%%%%%%%%%%%%%%%%%%%%%%%%%%%%%%%%%%%%%%%%%%%%%%%%%%%%
In this paper, we have presented the exact nonperturbative study of
strong correlation effects in topological insulators. We showed that the
clusters that can be studied with the Lanczos algorithm are sufficiently large
to identify a topological insulator, clearly distinguishing that the edge
states are conducting and the bulk states are insulating. For spinless
fermions in the Haldane model with repulsive nearest-neighbor interactions, we
found that for weak interaction strengths the system is a topological
insulator with circulating edge currents. For large interaction strengths, the
system is a topologically trivial CDW insulator. By using the fidelity
metric and the CDW structure factor, we showed that the choice of the cluster
is significant and that clusters must have a reciprocal space that
incorporates the zone corner of the Brillouin zone as a valid $k$ point. For
these clusters, we have observed that the transition from the topological
insulator to the topologically trivial CDW insulator is first order.

We also investigated the properties of the Haldane model with hard-core bosons
at half filling and $\phi = \pi / 4$. Here we found that the bosons are in a
superfluid phase for weak interaction strengths and a Mott insulator for large
interaction strengths. The growth of the fidelity metric with system size is
consistent with a second-order phase transition but scaling of the CDW
structure factor cannot distinguish whether the universality class, if any, is
Ising or XY. Additionally, no signature of topological order is present in
this system. However, this model exhibits interesting critical behavior with a
phase diagram that is currently being investigated.\cite{varney2010}

\begin{figure}[t]
  \includegraphics*[width=\columnwidth,viewport=0 198 612 605]{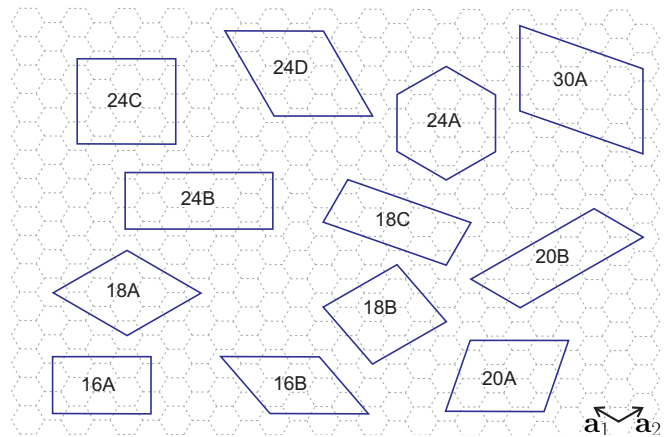}
  \caption{\label{fig:clusters}
    (Color online) Illustration of different clusters. Only the 18A, 18C, 24C,
    24D, and 30A clusters contain the zone corner as a valid $k$ point. The 24A
    cluster was studied with both open and periodic boundary conditions. The
    basis vectors for the lattice are shown in the bottom right of the
    figure. 
  }
\end{figure}

We note there that the interacting Haldane model by no means exhausts all
possible types of interacting lattice Hamiltonians, which may show interesting
interplay of strong correlations and topological textures, but it rather
represents the simplest case where the asymptotic regimes are well
understood. Therefore, it is conceivable that other models may host even more
interesting transitions and phases such as topological Mott insulators, which
would represent the lattice analogs of fractional quantum Hall states. Some
candidate states of this sort were previously discussed in
Refs.~\onlinecite{lee2004,seidel2005,burkov2010}, however no convincing
evidence for their existence has been found in any model.  We conclude by
pointing out that our work has important consequences for the studies of
topological insulating states in interacting lattice systems because it
demonstrates explicitly that most hallmark features of topological insulators
survive in small systems, which can be analyzed using unbiased exact
diagonalization methods and could be realized in experiments with trapped ions
or ultracold gases in optical lattices.\cite{bloch2008}

%%%%%%%%%%%%%%%%%%%%%%%%%%%%%%%%%%%%%%%%%%%%%%%%%%%%%%%%%%%%%%%%%%%%%%%%
%\section*{\label{sec:acknowledgments}ACKNOWLEDGMENTS}
%%%%%%%%%%%%%%%%%%%%%%%%%%%%%%%%%%%%%%%%%%%%%%%%%%%%%%%%%%%%%%%%%%%%%%%%
\begin{acknowledgments}
\label{sec:acknowledgments}
This research was supported by NSF through JQI-PFC (C.N.V., K.S., and V.G.),
U.S. Office of Naval Research (C.N.V. and M.R.), and US-ARO
(V.G.). C.N.V. thanks C.-C. J. Wang, E. Khatami, and K. Mikelsons for useful
input.
\end{acknowledgments}

\appendix*
\section{\label{appendix}Cluster Selection}
It is common in the Lanczos literature to diagonalize clusters with $L \times
L$ sites in addition to other square clusters which completely cover the
lattice.\cite{betts1996} For a honeycomb lattice, we generally follow a
similar prescription of choosing a parallelogram or ``tilted rectangle'' to
describe the cluster. These vectors can be written in terms of the basis
vectors ${\bf a}_1$ and ${\bf a}_2$, shown in Fig.~\ref{fig:clusters}, that
describe the underlying lattice
\begin{align}
  \begin{split}
    {\bf A}_1 &= m_1 {\bf a}_1 + m_2 {\bf a}_2,\\
    {\bf A}_2 &= n_1 {\bf a}_1 + n_2 {\bf a}_2,
  \end{split}
\end{align}
where $m_1$, $m_2$, $n_1$, and $n_2$ are integers. In Fig.~\ref{fig:clusters},
we explicitly show the shape of the clusters used in this work. Note that some
of these clusters do {\em not} have all of the symmetry properties of the
bulk. The lone exception to choosing a parallelogram to tile the lattice is
the cluster designated $24A$. This hexagon-shaped cluster can tile the lattice
and retains its complete symmetry. It is the only cluster studied with open
boundary conditions.

As mentioned in the main text above, it is crucial that one distinguish
between two different types of clusters, which depend on whether $K = (2\pi/3,
2\pi/\sqrt{3})$ is a valid point in reciprocal space. For this purpose, we
provide a necessary and sufficient condition for the $K$ point to be a valid
momentum point,
\begin{align}
  \label{eqn:condition}
  \begin{split}
    \alpha_1 &= 2 m_1 / 3 - m_2 / 3 \in \mathbb{Z},\\ 
    \alpha_2 &= 2n_1 / 3 - n_2 /3 \in \mathbb{Z}.
  \end{split}
\end{align}
If both $\alpha_1$ and $\alpha_2$ for a particular cluster are integers, then
$K$ is a valid point in the reciprocal space for that cluster. The number of
sites in a cluster is $N = 2|m_1 n_2 - m_2 n_1|$ and it can be easily checked
this condition can only be satisfied when the number of sites are $6 l$, where
$l$ is a positive integer. For the clusters we studied, only clusters of $N =
18, 24$, and $30$ sites can contain $K$ point as a momentum point. Notice that
a different choice of ${\bf a}_1$ and ${\bf a}_2$ will change the condition
described in Eq.~\eqref{eqn:condition}.

\bibliography{references}

\end{document}